\documentclass{ws-procs9x6}
\usepackage{subfigure}

\begin{document}

\hyphenation{dis-pla-ce-ment re-si-sti-vi-ty ge-ne-ra-ted re-la-ti-vi-stic dif-fe-ren-ce in-sensi-ti-ve ki-ne-ma-ti-cal Sy-ste-ma-tic sy-ste-ma-tic ave-ra-ging con-ver-ters con-ver-ter la-bo-ra-to-ry se-con-da-ry
se-con-da-ries si-mu-la-tion si-mu-la-tions do-mi-na-te do-mi-na-tes ta-bu-la-tions
Ta-bu-la-tions Di-stri-bu-tions
di-stri-bu-tions di-stri-bu-tion Di-stri-bu-tion Di-spla-ce-ment di-spla-ce-ment Di-spla-ce-ments
di-spla-ce-ments Li-near li-near Ca-sca-de Ca-sca-des ca-sca-de ca-sca-des
ta-bu-la-tion Ta-bu-la-tion re-pe-ti-ti-ve E-lec-tron E-lec-trons
e-lec-tron e-lec-trons De-tec-tion Pro-duc-tion pro-duc-tion
Re-so-lu-tions Re-so-lu-tion re-so-lu-tions re-so-lu-tion
Ope-ra-tion mi-ni-mum Ener-gy Ener-gies fer-ro-mag-net
fer-ro-mag-nets meta-sta-ble meta-sta-bi-lity con-fi-gu-ra-tion
con-fi-gu-ra-tions expo-nen-tially mo-bi-li-ty- mo-bi-li-ties
tem-pe-ra-tu-re tem-pe-ra-tu-res con-cen-tra-tion con-cen-tra-tions
elec-tro-nic elec-tro-nics STMelec-tro-nics sec-tion Sec-tion
Chap-ter chap-ter theo-ry ap-pro-xi-mation ra-dia-tion Ra-dia-tion
ca-pa-ci-tan-ce approaches tran-sport dispersion Ca-lo-ri-me-try
ca-lo-ri-me-try En-vi-ron-ment En-vi-ron-ments en-vi-ron-ment
en-vi-ron-ments Fur-ther-mo-re do-mi-nant ioni-zing pa-ra-me-ter pa-ra-me-ters
O-sa-ka ge-ne-ral exam-ple Exam-ple ca-vi-ty Ca-vi-ty He-lio-sphe-re
he-lio-sphe-re dis-tan-ce Inter-pla-ne-ta-ry inter-pla-ne-ta-ry
ge-ne-ra-li-zed sol-ving pho-to-sphe-re sym-me-tric du-ring
he-lio-gra-phic strea-ming me-cha-nism me-cha-nisms expe-ri-mental
Expe-ri-mental im-me-dia-tely ro-ta-ting na-tu-rally
ir-re-gu-la-ri-ties o-ri-gi-nal con-si-de-red e-li-mi-na-ting
ne-gli-gi-ble stu-died dif-fe-ren-tial mo-du-la-tion ex-pe-ri-ments
ex-pe-ri-ment Ex-pe-ri-ment Phy-si-cal phy-si-cal in-ve-sti-ga-ted
Ano-de Ano-des ano-de ano-des re-fe-ren-ce re-fe-ren-ces
ap-pro-xi-ma-ted ap-pro-xi-ma-te in-co-ming bio-lo-gi-cal
atte-nua-tion other others eva-lua-ted nu-cleon nu-cleons reac-tion
pseu-do-ra-pi-di-ty pseu-do-ra-pi-di-ties esti-ma-ted va-lue va-lues
ac-ti-vi-ty ac-ti-vi-ties bet-ween Bet-ween dis-cre-pan-cy
dis-cre-pan-cies cha-rac-te-ri-stic
cha-rac-te-ri-stics sphe-ri-cally anti-sym-metric ener-gy ener-gies
ri-gi-di-ty ri-gi-di-ties leaving pre-do-mi-nantly dif-fe-rent
po-pu-la-ting acce-le-ra-ted respec-ti-ve-ly sur-roun-ding pa-ral-lel
sa-tu-ra-tion vol-tage vol-tages da-ma-ge da-ma-ges be-ha-vior
equi-va-lent si-li-con exhi-bit exhi-bits con-duc-ti-vi-ty
con-duc-ti-vi-ties dy-no-de dy-no-des created Fi-gu-re Fi-gu-res
tran-si-stor tran-si-stors Tran-si-stor Tran-si-stors ioni-za-tion
Ioni-za-tion ini-tia-ted sup-pres-sing in-clu-ding maxi-mum mi-ni-mum
vo-lu-me vo-lu-mes tu-ning ple-xi-glas using de-pen-ding re-si-dual har-de-ning li-quid
know-ledge usage me-di-cal par-ti-cu-lar scat-te-ring ca-me-ra se-cond hea-vier hea-vy trans-axial
con-si-de-ration created Hy-po-the-sis hy-po-the-sis usually inte-ra-ction Inte-ra-ction
inte-ra-ctions Inte-ra-ctions pro-ba-bi-li-ty pro-ba-bi-li-ties
fol-low-ing cor-re-spon-ding e-la-stic readers reader pe-riod pe-riods geo-mag-ne-tic sa-ti-sfac-tory}

\begin{center}
To appear on the Proceedings of the 13th ICATPP Conference on\\
Astroparticle, Particle, Space Physics and Detectors\\ for Physics Applications,\\ Villa  Olmo (Como, Italy), 3--7 October, 2011, \\to be published by World Scientific (Singapore).
\end{center}
\vspace{-1.5cm}

\title{ELECTRICAL CHARACTERIZATION OF SiPM AS A FUNCTION OF TEST FREQUENCY AND TEMPERATURE}

\author{
M.J. Boschini$^{1,3}$, C. Consolandi$^{1}$, P.G. Fallica$^{4}$, M. Gervasi$^{1,2}$, D. Grandi$^{1}$, \\M. Mazzillo$^{4}$, S. Pensotti$^{1,2}$, P.G. Rancoita$^{1}$, D. Sanfilippo$^{4}$, \\M. Tacconi$^{1}$* and G. Valvo$^{4}$
}

\address{$^{1}$Istituto Nazionale di Fisica Nucleare, INFN Milano-Bicocca, Milano (Italy) \\
$^{2}$Department of Physics, University of Milano Bicocca, Milano (Italy) \\
$^{3}$CILEA, Segrate (MI) (Italy) \\
$^{4}$STMicroelectronics, Catania (Italy)
}
\address{\textbf{E-mail: mauro.tacconi@mib.infn.it}}

\begin{abstract}
Silicon Photomultipliers (SiPM) represent a promising alternative to classical photomultipliers for the detection of photons in high energy physics and medical physics, for instance.~In the present work, electrical characterizations of test devices - manufactured by STMicroelectronics - are presented.~SiPMs with an area of $3.5 \times 3.5\,\textrm{mm}^2$ and a cell pitch of $54\,\mu$m were manufactured as arrays of $64 \times 64$ cells and exhibiting a fill factor of 31\%.~The capacitance of SiPMs was measured as a function of reverse bias voltage at frequencies ranging from about 20\,Hz up to 1\,MHz and temperatures from 310\,K down to 100\,K.~Leakage currents were measured at temperatures from 410\,K down to 100\,K.~Thus, the threshold voltage - i.e., the voltage above a SiPM begins to operate in Geiger mode - could be determined as a function of temperature.~Finally, an electrical model capable of reproducing the frequency dependence of the device admittance is presented.
\end{abstract}
\section{Introduction}
In recent years Silicon Photomultipliers (SiPM) has been developed for usage in photon detection.~The high gain, insensitivity to magnetic field and low reverse bias voltage of operation make SiPM a promising candidates as replacement of classical photomultipliers\cite{intro} in several of their applications.~The SiPM is an array of parallel-connected single photon avalanche diodes (SPAD)\cite{Maz1}\,.~Every SPAD operates in Geiger mode with a quenching resistor in series to prevent an avalanche multiplication process from taking place.~The overall output of a SiPM depends on how many SPADs are simultaneously ignited\cite{intro1}$^{,}$\cite{Maz2}\,.~
\par
 The current SiPM device was manufactured by STMicroelectronics (details of the technology can be found in Refs.\cite{Maz1,Condorelli}).~It consisted of an array of $64 \times 64$ SPAD cells with $3.5 \times 3.5\,\textrm{mm}^2$ effective area, a cell pitch of 54$\,\mu$m and a fill factor of $\approx 31$\%.~
 \par
 In the present article, the electrical characteristics of these devices are shown as function of test frequency from 1\,MHz down to about 20\,Hz and temperature from 410\,K down to 100\,K (Sects.~\ref{Elec_Cha_SiPM}--\ref{Cur_V_SiPM}).~In addition, in Sects.~\ref{Elec_Mod} and~\ref{Elec_Mod_SiPM} an electrical model is discussed and compared with data obtained from dependencies of capacitance and resistance on the test frequency for both a photodiode and a SiPM devices.
%%%%%%%%%%%%%%%%%%%%%%%%%%%%%%%%%%%%%%%%%%%%%%%%%%%%%%%%%
\begin{figure}[b]
\begin{center}
\vskip -.3cm
 \includegraphics[width=0.8\textwidth]{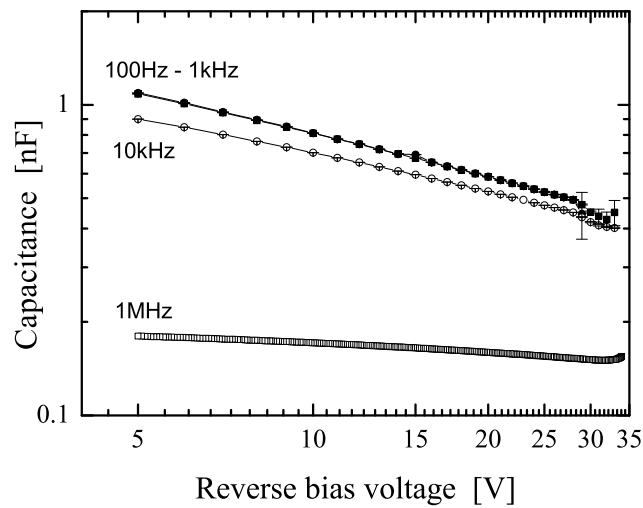}
 \vskip -.3cm
 \caption{Capacitance (in nF) as a function of reversed bias voltage (in V) using the LCZ meter with test frequencies of 100\,Hz ($\bullet$), 1\,kHz ($\blacksquare$), 10\,kHz ($\circ$) and BC with 1\,MHz ($\square$).}\label{cv1}
\end{center}
\end{figure}
\begin{figure}[t]
\begin{center}
%\vskip -.3cm
 \includegraphics[width=0.8\textwidth]{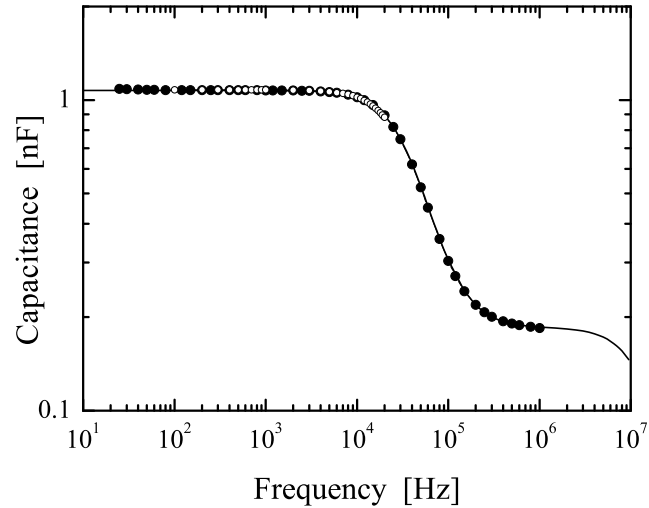}
 \vskip -.3cm
 \caption{Capacitance (in nF) as a function of the frequency (in Hz) for a sample operated with a reverse bias of 5\,V: data points indicated with $\bullet$ ($\circ$) were obtained using the LCR (LCZ) meter, the continuous line is obtained from Eq.~(\ref{susceptance_array}) (see discussion in Sect.~\ref{Elec_Mod_SiPM}).}\label{cv2}
\end{center}
\end{figure}
%%%%%%%%%%%%%%%%%%%%%%%%%%%%%%%%%%%%%%%%%%%%%%%%%%%%%%%%%%%%%%%%%%%%%%%%%%%%%%%%%%%%%%%%%%%%%%%%%%%%%%%%%%%%%%%%%%%
\begin{figure}[t]
\begin{center}
\vskip -.3cm
 \includegraphics[width=1.0\textwidth]{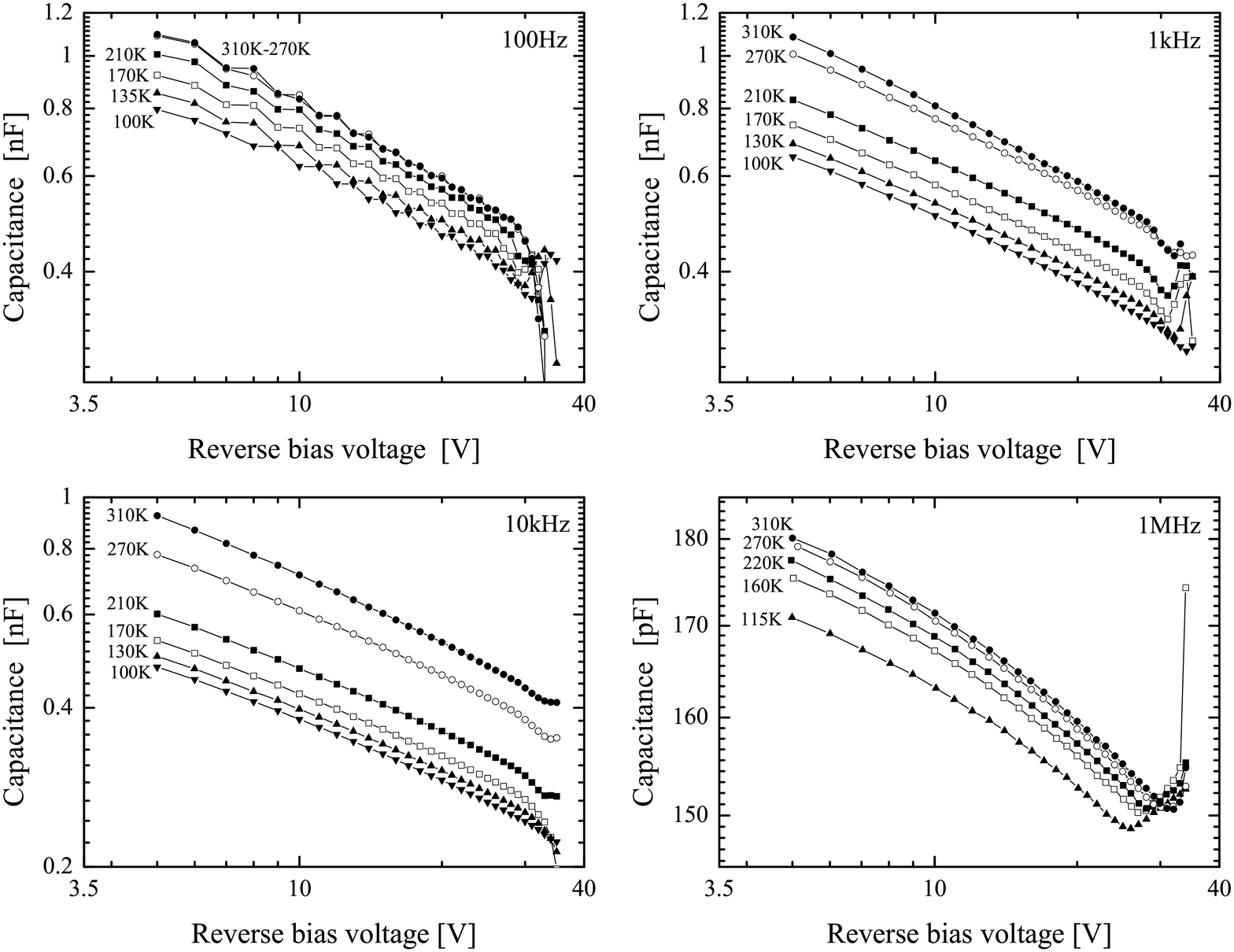}
 \vskip -.3cm
 \caption{Capacitance response (in nF or pF) obtained using the LCZ meter as a function of reverse bias applied (in V) and temperature from 310\,K down to 100\,K: 100\,Hz (upper left), 1\,kHz (upper right), 10\,kHz (bottom left) and - using the BC - 1\,MHz (bottom right).}\label{cvt}
\end{center}
\end{figure}
%%%%%%%%%%%%%%%%%%%%%%%%%%%%%%%%%%%%%%%%%%%%%%%%%%%%
%%%%%
\section{Electrical Characteristics of SiPM Devices}
\label{Elec_Cha_SiPM}
The electrical characteristics of SiPM devices were investigated as function of applied reverse bias voltage ($V_{\rm r}$) and temperature.~The capacitance response was also studied as a function of test frequency of the capacimeter employed for such a measurement (Sect.~\ref{C_resp_SiPM}).~
\par
Furthermore (Sect.~\ref{Cur_V_SiPM}), the measurements of the SiPM leakage current allowed one to determine (as a function of temperature) the value (and its temperature dependence) of the so-called \textit{threshold voltage} ($V_{\rm th}$),~i.e., the reverse bias voltage above which a SiPM begins to operate in Geiger mode.
%%%%%%%%%%%%%%%%%%%%%%%%%%
\begin{figure}[b]
\begin{center}
%\vskip -.7cm
 \includegraphics[width=1.0\textwidth]{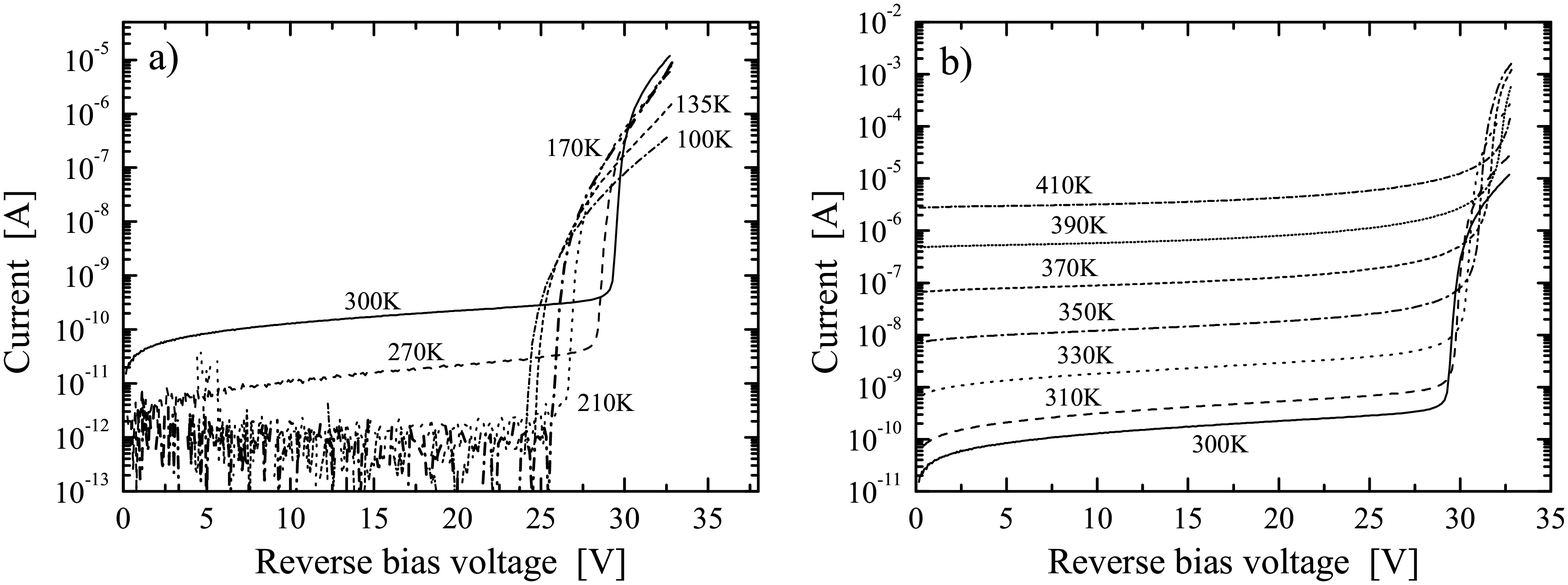}
 \vskip -.3cm
 \caption{(a) Leakage current (in A) as a function of reverse bias voltage (in V) from 310 down to 100\,K; (b) leakage current (in A) as a function of reverse bias voltage (in V) from 300 up to 410\,K. }\label{ivt}
\end{center}
\end{figure}
%%%%%%%%%%%%%%%%%%%%%%%%%%%%
%
\subsection{Capacitance Response}
\label{C_resp_SiPM}
The capacitance response of SiPM devices was systematically measured as a function of reverse bias voltage applied and test frequency at 300\,K (i.e., at room temperature).~Moreover, using a liquid nitrogen cryostat, at a few fixed frequencies the capacitance was also measured as a function of reverse bias voltage and temperature from 310\,K down to 100\,K.~For these purposes three instruments were employed,~i.e., a Boonton capacimeter (BC) (using its internally provided power source and a test frequency of 1\,MHz), an LCZ (Agilent Technologies 4276A) and LCR (Agilent Technologies 4284A) meters with the reverse bias supplied to the device from an external power supply.~The LCZ meter employed test frequencies from 100\,Hz up to 20\,KHz; the LCR meter from $\approx 20\,$Hz up to 1\,MHz.~Furthermore, the measurements were performed selecting the parallel equivalent-circuit mode of the LCR and LCZ meters for the device under test (DUT).~Thus, the device response is assumed to be that one from a parallel capacitor-resistor circuit (see discussion in Sect.~\ref{Elec_Mod}).~Finally, the currently reported experimental errors are the standard deviations obtained using 50 subsequent measurements of the device capacitance under the same experimental conditions.
\par
In Fig.~\ref{cv1}, the capacitance response of a SiPM using frequencies of 100\,Hz, 1\,kHz, 10\,kHz and 1\,MHz is shown as a function of $V_{\rm r}$ at 300\,K.~Above 1\,kHz, the measured capacitance decreases with the increase of the test frequency.~In Fig.~\ref{cv2}, the capacitance response - determined using both LCZ and LCR meters - is shown for a device operated with a reverse bias voltage of 5\,V at 300\,K.~It can be observed that above $\approx 10$\,kHz the measured capacitance largely decreases with the increase of the test frequency of the instrument up to achieving an almost constant response above a few hundreds of kHz.
\par
In Fig.~\ref{cvt}, the capacitance response using 100\,Hz, 1\,kHz, 10\,kHz and 1\,MHz is shown as a function of $V_{\rm r}$ (in V) and temperature from 310\,K down to 100\,K.~The measured capacitance is observed to decrease with decreasing temperature.~This behavior is also observed for frequencies lower than $\approx 10\,$kHz; while a similar capacitance response was exhibited at 300\,K (see Fig.~\ref{cv1}).
\subsection{Current-Voltage Characteristics}
\label{Cur_V_SiPM}
%

%%%%%%%%%%%%%%%%%%%%%%%%%%%%%%%
\begin{figure}[b]
\begin{center}
\vskip -.3cm
 \includegraphics[width=1.0\textwidth]{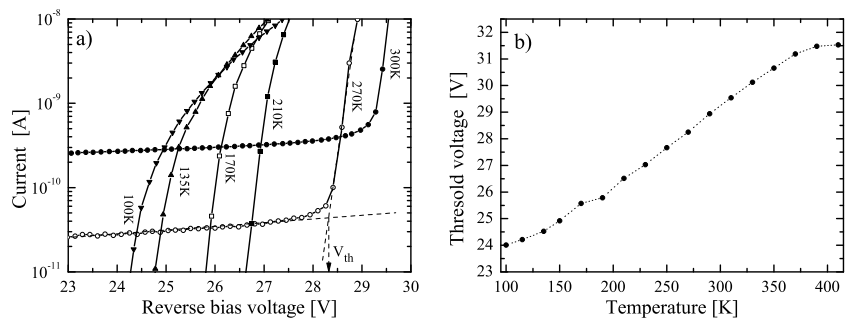}
 \vskip -.3cm
 \caption{(a) Leakage current (in A) as a function of reverse bias voltage applied (in V) and an example of threshold voltage, $V_{\rm th}$ in V, determination at 270\,K; (b) threshold voltage (in V) as a function of temperature (in K).}\label{ivt1}
\end{center}
\end{figure}
%%%%%%%%%%%%%%%%%%%%%%%%%%%%%
The dependence of leakage current on reverse bias was studied as a function of temperature from 410\,K down to 100\,K as shown in Fig.~\ref{ivt}.~With increasing $V_{\rm r}$ the leakage current does not vary by more than an order of magnitude until the multiplication regime is ignited,~i.e., so far the \textit{threshold voltage}, $V_{\rm th}$, is reached.~$V_{\rm th}$ is the voltage corresponding to the value of leakage current at the intercept between the two $I$ versus $V$ curves obtained, the first when the multiplication regime is sharply starting and the second one at lower voltages,~i.e., before the multiplication occurs: an example regarding the determination of the value of $V_{\rm th}$ at 270\,K is reported in Fig.~\ref{ivt1}(a).~As shown in Fig.~\ref{ivt1}(b), the threshold voltage lowers -~thus, SPADs turn into a Geiger-mode regime at progressively lower reverse bias voltages~- with lowering temperature at the rate of $\approx - 29 \,$mV/K from $\approx 360$ down to $\approx 130$\,K.
%%%%%%%%%%%%%%%%%%%%%%%%%%
\begin{figure}[t]
\begin{center}
\vskip -.3cm
 \includegraphics[width=1.0\textwidth]{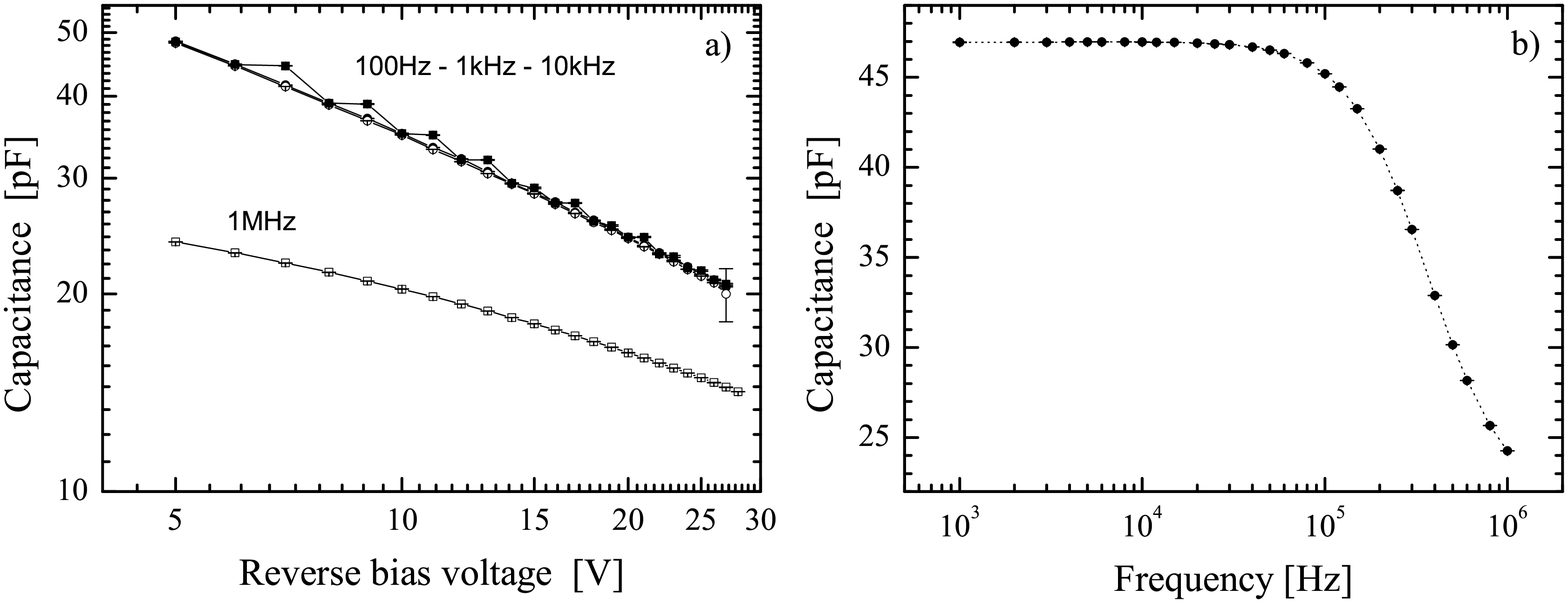}
 \vskip -.3cm
 \caption{(a) Capacitance (in pF) of a photodiode as a function of reverse bias voltage (in V) at 100\,Hz, 1\,kHz, 10\,kHz (using the LCZ meter) and 1\,MHz (using the BC); (b) capacitance (in pF) of a photodiode as a function of test frequency  in Hz (using the LCR meter) with an applied reverse biased voltage of 6\,V.}\label{cf1}
\end{center}
\end{figure}
%%%%%%%%%%%%%%%%
%\vspace{-.3cm}
%
\section{Electrical Model}
\label{Elec_Mod}
At room temperature, the electrical frequency response of SiPMs was further investigated using a test device manufactured by STMicroelectronics.~The latter device was a photodiode (PD) with a structure similar to that of a SPAD cell, an active area of $\approx 0.2\,$mm$^{2}$, but no quenching resistor in series.~In Fig.~\ref{cf1}(a), the capacitance of the PD is shown as a function of reverse bias voltage at 100\,Hz, 1\,kHz, 10\,kHz and 1\,MHz.~Furthermore, one can see that the device exhibits almost no dependence on the test frequency below $\approx 100\,$kHz [Fig.~\ref{cf1}(b)].
%%%%%%%%%%%%%%%%%%%%%%%%%%%%%%%%%%%
\begin{figure}[b]
\begin{center}
\vskip -.1cm
 \includegraphics[width=0.7\textwidth]{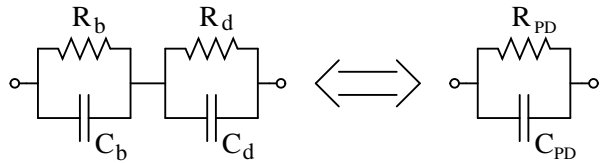}
 \vskip -.3cm
 \caption{SIJD model for silicon photodiodes and radiation detectors: $C_{\rm d}$ ($C_{\rm b}$) and $R_{\rm d}$ ($R_{\rm b}$) are the capacitance and resistance of the depleted (field free) region, respectively; $C_{\rm PD}$ and $R_{\rm PD}$ are the overall capacitance and resistance of the photodiode, respectively.}\label{model}
\end{center}
\end{figure}
%%%%%%%%%%%%%%%%%%%%%%%%
%%%%%%%%%%%%%%%%%%%%%%%%%%
\begin{figure}[t]
\begin{center}
%\vskip -.1cm
 \includegraphics[angle=0,width=.75\textwidth]{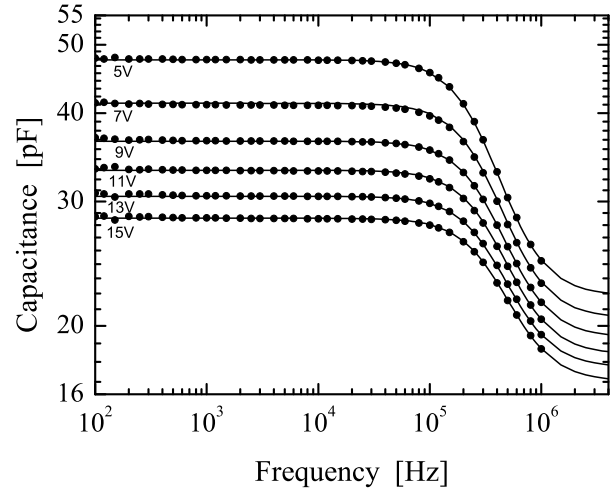}
 \vskip -.3cm
 \caption{Measured capacitance ($\bullet$), $C_{\rm PD}$, (in pF) as a function of test frequency (in Hz) with superimposed continuous lines obtained from the SIJD model [Eq.~(\ref{susceptance_PD})]. }\label{cfmodel}
\end{center}
%%%%%%%%%%%%%%%%%%%%%%
\end{figure}
\begin{figure}[b]
\vskip -.1cm
\begin{center}
 \includegraphics[angle=0,width=.75\textwidth]{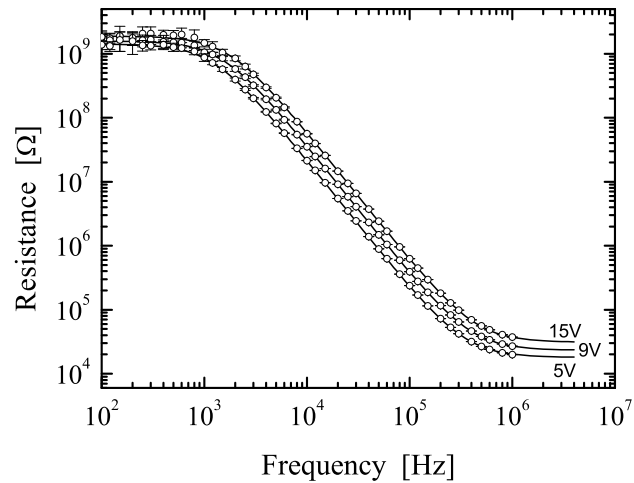}
 \vskip -.3cm
 \caption{Measured resistance ($\circ$), $R_{\rm PD}$, (in $\Omega$) as a function of test frequency (in Hz) at 5, 9 and 15\,V with superimposed continuous lines obtained from the SIJD model [Eq.~(\ref{conductance_PD})]. }\label{rfmodel}
\end{center}
\end{figure}
%%%%%%%%%%%%%%%%%
%%%%%%%%%%
\begin{figure}[t]
\vskip -.3cm
\begin{center}
 \includegraphics[width=1.0\textwidth]{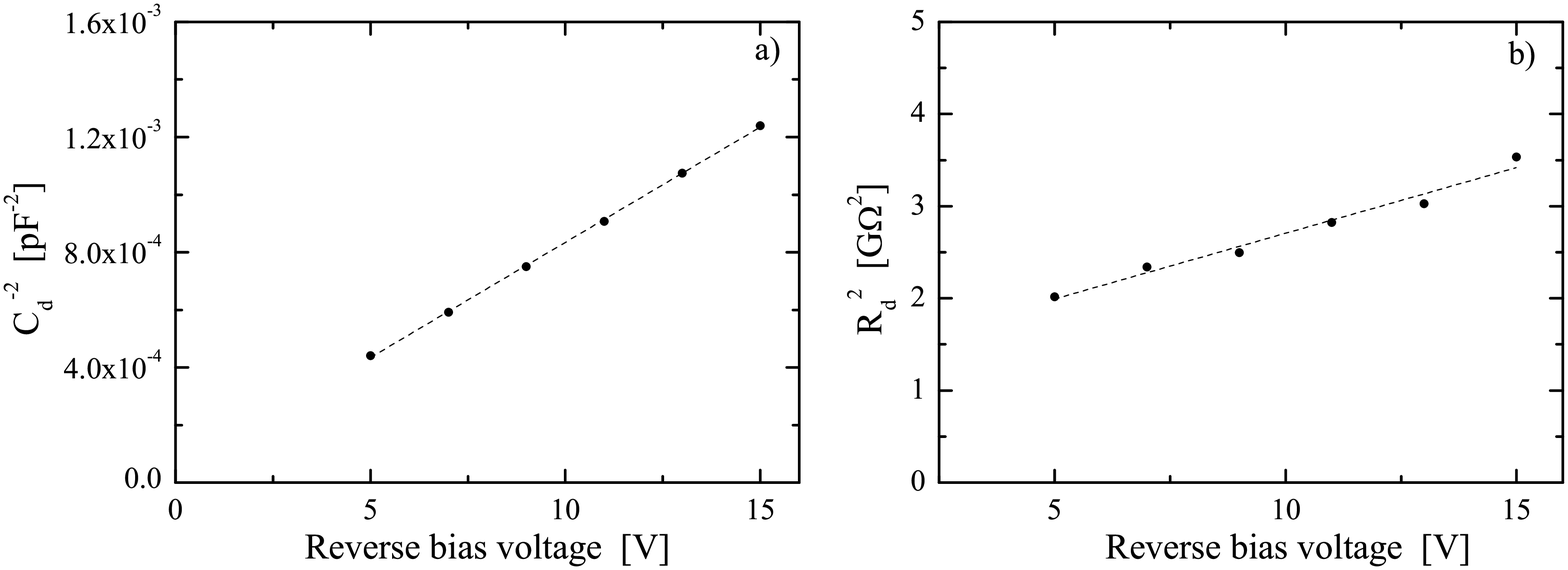}
 \vskip -.3cm
 \caption{(a) $1/C_{\rm d}^2$ (in pF$^{-2}$) and (b) $R_{\rm d}^2$ (in G$\Omega^{2}$) from Table~\ref{Conclusion} as a function of applied reversed bias voltage in V.~The dashed lines are those from a linear fit to the data.}\label{cfval}
\end{center}
\end{figure}
%%%%%%%%%%%%%%%%%%%%%%%%%%%%%%%%%%%%%%%%%%%%%%%%
\par
For silicon photodiodes and radiation detectors, the electrical response down to cryogenics temperatures is usually modeled using the so-called \textit{small-signal ac impedance of junction diode} (SIJD) operated under reverse bias (e.g., see Refs.\cite{Champ,rop_si}\,, Section~4.3.4 of~Ref.\cite{LR_3rd} and references therein).~In the framework of the SIJD model, the device consists of the depleted and field free regions connected in series.~Each one of these regions, in turn, consists of a parallel-connected capacitor and resistor.~In Fig.~\ref{model}, the equivalent circuit of a photodiode is shown: $C_{\rm d}$ ($C_{\rm b}$) and $R_{\rm d}$ ($R_{\rm b}$) are the capacitance and resistance of the depleted (field free) region, respectively.~
\par
Labeling (Fig.~\ref{model}) the overall\footnote[4]{$C_{\rm PD}$ and $R_{\rm PD}$ are the quantities directly measured selecting the parallel equivalent-circuit mode of the LCR and LCZ meters.} capacitance and resistance of a photodiode with $C_{\rm PD}$ and $R_{\rm PD}$, respectively, one can express the admittance of the device as:
\begin{eqnarray}
\nonumber Y_{\rm PD}(\omega) &=&  G_{\rm PD}(\omega)+ j\,B_{\rm PD}(\omega)\\
\label{admittance_PD} &=& \frac{1}{R_{\rm PD}} +
j \, \omega C_{\rm PD}
\end{eqnarray}
with $\omega= 2\pi f$, where $f$ is the test frequency. The conductance (i.e., the real part of the admittance) and susceptance (the imaginary part of the admittance) are respectively given by:
\begin{eqnarray}
\label{conductance_PD}  G_{\rm PD}(\omega) = \frac{1}{R_{\rm PD}} &=& \frac{R_{\rm d}+R_{\rm b}+\omega^{2}R_{\rm d}R_{\rm
b}(R_{\rm d}C_{\rm d}^{2}+R_{\rm b}C_{\rm b}^{2})} {(R_{\rm
d}+R_{\rm b})^{2}+\omega^{2}R_{\rm d}^{2}R_{\rm b}^{2}(C_{\rm
b}+C_{\rm d})^{2}} \\
\label{susceptance_PD}  B_{\rm PD}(\omega) = \omega C_{\rm PD} &=&  \omega \left[ \frac{\omega^{2}R_{\rm d}^{2}R_{\rm
b}^{2}C_{\rm d}C_{\rm b}(C_{\rm b}+C_{\rm d})+ R_{\rm d}^{2}C_{\rm
d}+R_{\rm b}^{2}C_{\rm b}}{(R_{\rm d}+R_{\rm
b})^{2}+\omega^{2}R_{\rm d}^{2}R_{\rm b}^{2}(C_{\rm b}+C_{\rm
d})^{2}}\right]
\end{eqnarray}
[e.g.,~see Equations~(4.184, 4.185) at page~455 of~Ref.\cite{LR_3rd}].~
\par
It has to be remarked that the values of both $C_{\rm PD}$ and $R_{\rm PD}$ are those which can be determined, for instance, selecting the parallel equivalent-circuit mode for the DUT using the LCR meter.~In the SIJD model, the values of $C_{\rm d}$, $R_{\rm d}$, $C_{\rm b}$ and $R_{\rm b}$ do not depend on the test frequency.~In addition, $C_{\rm d}$, $R_{\rm d}$ are expected to depend on the applied reverse bias voltage, because the depleted layer width increases with increasing $V_{\rm r}$.~
\par
$C_{\rm d}$, $R_{\rm d}$, $C_{\rm b}$ and $R_{\rm b}$ were determined (Table~\ref{Conclusion}) as a function $V_{\rm r}$, by  a fit to the measured quantities $C_{\rm PD}$  and $R_{\rm PD}$ (obtained using the LCR meter) using the corresponding expressions [e.g.,~see Eqs.~(\ref{conductance_PD},~\ref{susceptance_PD})] obtained from the SIJD model for a photodiode.~For instance, in Fig.~\ref{cfmodel} (Fig.~\ref{rfmodel}) the experimental data and fitted curves are shown for the capacitance (resistance) measurements as a function of test frequency and applied reverse bias voltage.~As expected\cite{Champ}\, (Table~\ref{Conclusion}), the field free region is almost independent of $V_{\rm r}$, while both the capacitance and resistance of the depleted region exhibit a dependence on the reverse bias.~In Fig.~\ref{cfval}, $1/C_{\rm d}^2$ [Fig.~\ref{cfval}(a)] and $R_{\rm d}^2$ [Fig.~\ref{cfval}(b)] values from Table~\ref{Conclusion} are shown as a function of applied reverse voltage.~Although the junction photodiode cannot be considered as a one-sided step junction\footnote[5]{The depletion layer characteristics as a function of reverse bias voltage can be found treated, for instance, in Chapter~6-2 of Ref.\cite{Wolf} (see also Ref.\cite{Fair}).}, the the dashed curves (Fig.~\ref{cfval}) obtained from fits to the reported data -~assuming a linear dependence of $1/C_{\rm d}^2$ and $R_{\rm d}^2$ on $V_{\rm r}$~- are well suited for reproducing the dependence on the reverse bias voltage.
%%%%%%%%%%%%%
\begin{table}[t]
\begin{center}
\tbl{$C_{\rm d}$, $R_{\rm d}$, $C_{\rm b}$ and $R_{\rm b}$ obtained from a fit of the SIJD model for a photodiode to experimental data as a function of reverse bias voltage ($V_{\rm r}$).}
{\begin{tabular}{rl|ccccc}
\toprule
$V_{\rm r}$ [V]& & {$C_{\rm d}$ [pF]}  & {$R_{\rm d}$ [G$\Omega$]}  & {$C_{\rm b}$ [pF]}  &{$R_{\rm b} $[k$\Omega$]}  \\
\colrule
5 & & 47.6 & 1.42 & 41.1  & 5.2  \\
7 & & 41.1 & 1.53 & 40.8 & 5.2 \\
9 & & 36.5 & 1.58 & 40.8 & 5.2 \\
11 & & 33.2 & 1.68 & 40.4 & 5.2 \\
13 & & 30.5 & 1.74 & 40.9 & 5.3 \\
15 & & 28.4 & 1.88 & 40.5 & 5.3 \\
\botrule
\end{tabular}}\label{Conclusion}
\end{center}
\end{table}
%%%%%%%%%%%%%%%%%%%%%%%%
\vspace{-.3cm}
\subsection{Electrical Model for SiPMs}
\label{Elec_Mod_SiPM}
\begin{figure}[t]
\vskip -.3cm
\begin{center}
 \includegraphics[width=0.75\textwidth]{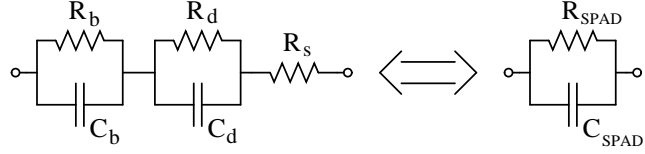}
 \vskip -.3cm
 \caption{SIJD model for a SPAD cell: $C_{\rm d}$ ($C_{\rm b}$) and $R_{\rm d}$ ($R_{\rm b}$) are, respectively, the capacitance and resistance of the depleted (field free) region with a series resistance $R_{\rm s}$; $C_{\rm spad}$ and $R_{\rm spad}$ are the overall capacitance and resistance of a SPAD cell, respectively.}\label{model_spad}
\end{center}
\end{figure}

A SiPM device consists of a set of SPAD devices (4096 in the current SiPM under test) connected in parallel.~The equivalent electrical circuit of the elemental cell (i.e.,~a SPAD cell) is shown in Fig.~\ref{model_spad} and differs from that of a photodiode (discussed in Sect.~\ref{Elec_Mod}) by a resistance ($R_{\rm s}$) added in series.~In Fig.~\ref{model_spad},
$C_{\rm d}$ ($C_{\rm b}$) and $R_{\rm d}$ ($R_{\rm b}$) are the capacitance and resistance of the depleted (field free) region of the photodiode, respectively.~In the present technology of STMicroelectronics, $R_{\rm s}$ is typically about (0.2--1.0)\,M$\Omega$.
%%%%%%%%%%%%%%%%%%
%%%%%%%%%%%%%%%%%%%%%%
\begin{figure}[b]
\vskip -.3cm
\begin{center}
 \includegraphics[width=0.7\textwidth]{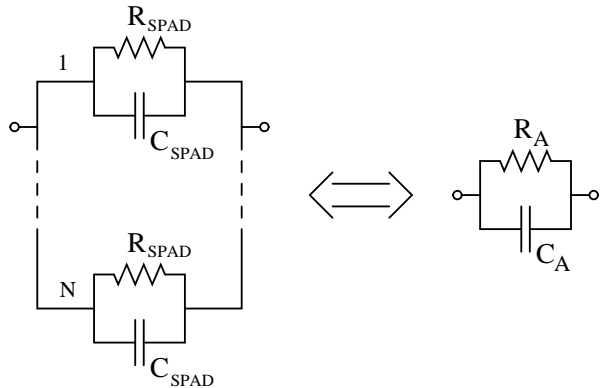}
 \vskip -.3cm
 \caption{SIJD model for a SiPM device consisting of N SPAD devices connected in parallel: $C_{\rm spad}$ and $R_{\rm spad}$ are, respectively, the capacitance and resistance of the equivalent circuit of a SPAD cell (Fig \ref{model_spad}), $C_{\rm A}$ and $R_{\rm A}$ those of the SiPM device.}\label{model_array}
\end{center}
\end{figure}
%%%%%%%%%%%%%%%%%%%%%
\par
The admittance ($Y_{\rm spad}$) of a SPAD cell is given by
\begin{eqnarray}
\nonumber  Y_{\rm spad}(\omega) &=&  G_{\rm spad}(\omega)+ j\,B_{\rm spad}(\omega)\\
\label{admittance_spad}  &=& \frac{1}{R_{\rm spad}} +
j \, \omega C_{\rm spad}
\end{eqnarray}
with $C_{\rm spad}$ and $R_{\rm spad}$ respectively expressed in terms of $C_{\rm d}$, $C_{\rm b}$, $R_{\rm d}$, $R_{\rm b}$ and $R_{\rm s}$ as
\begin{eqnarray}
\nonumber C_{\rm spad} \!&\!=\!&\!\frac{B_{\rm spad}(\omega)}{\omega}\\
\label{capacitance_spad}\! &\!=\! &\!\frac{\omega^{2}R_{\rm d}^{2}R_{\rm b}^{2}C_{\rm d}C_{\rm b}(C_{\rm b}+C_{\rm d})+ R_{\rm d}^{2}C_{\rm d}+R_{\rm b}^{2}C_{\rm b}}{D_{1}+D_{2}}, \\
\nonumber R_{\rm spad} \!&\!=\!&\!\frac{1}{G_{\rm spad}(\omega)}\\
\label{resistence_spad}\! &\!=\!& \!\frac{D_{1}\!+\!D_{2}}{R_{\rm b}\!+\!(1\!+\!\omega^{2}C_{\rm b}^{2}R_{\rm b}^{2})(R_{\rm s}\!+\!R_{\rm d})\!+\!\omega^{2}R_{\rm d}^{2}C_{\rm d}^{2}(\omega^{2}C_{\rm b}^{2}R_{\rm s}R_{\rm b}^{2}\!+\!R_{\rm s}\!+\!R_{\rm b})},
\end{eqnarray}
where $D_1$ and $D_2$ are
\begin{eqnarray}
\nonumber D_{1} &=& \omega^{2}R_{\rm b}^{2}C_{\rm b}^{2}(R_{\rm s}+R_{\rm d})^{2}+\omega^{2}R_{\rm d}^{2}C_{\rm d}^{2}(R_{\rm s}+R_{\rm b})^{2}\\
\nonumber D_{2} &=& (R_{\rm d}+R_{\rm b}+R_{\rm s})^{2}+\omega^{2}R_{\rm d}^{2}R_{\rm b}^{2}C_{\rm d}C_{\rm b}(2+\omega^{2}R_{\rm s}^{2}C_{\rm d}C_{\rm b}).
%N_{1} &=& R_{\rm b}+(1+\omega^{2}C_{\rm b}^{2}R_{\rm b}^{2})(R_{\rm s}+R_{\rm d})\\
%N_{2} &=& \omega^{2}R_{\rm d}^{2}C_{\rm d}^{2}(\omega^{2}C_{\rm b}^{2}R_{\rm s}R_{\rm b}^{2}+R_{\rm s}+R_{\rm b}).
\end{eqnarray}
%%%%%%%%%%%%%%%%%%%%%%%%
\begin{figure}[t]
\vskip -.3cm
\begin{center}
 \includegraphics[width=0.75\textwidth]{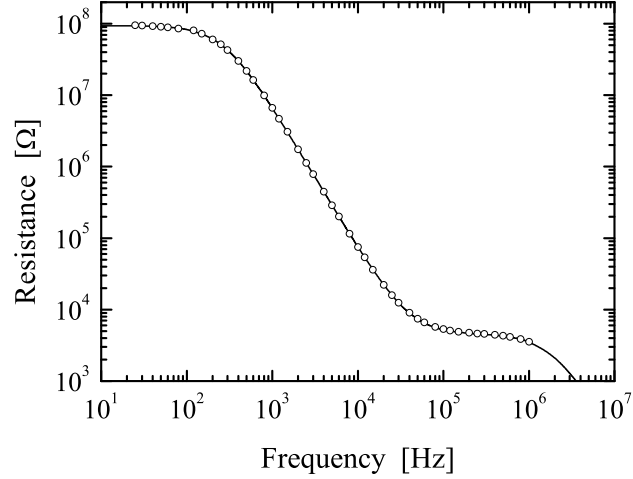}
 \vskip -.3cm
 \caption{Measured resistance ($\circ$), $R_{\rm A}$, (in $\Omega$) as a function of frequency (in Hz) with superimposed the continuous line obtained from Eq.~(\ref{conductance_array}).}\label{Array_R}
\end{center}
\end{figure}
%%%%%%%%%%%%%%%%%%%%%%%
\par
The equivalent electrical circuit\footnote[1]{It has to be remarked that one can also add a series resistance to $R_{\rm A}$.~This was the case for another SiPM device with 3600 SPAD cells manufactured by STMicroelectronics; for such a device a small additional series resistance of about 1\,k$\Omega$ was needed.} of a SiPM is shown in Fig.~\ref{model_array}.~Assuming that each SPAD cell has the same admittance, one finds that the overall admittance of an array is determined by:
\begin{eqnarray}
\nonumber Y_{\rm A}(\omega) &=& N \, Y_{\rm spad}(\omega)\\
\nonumber  &=& G_{\rm A}(\omega)+ j\,B_{\rm A}(\omega)\\
\label{admittance_array} &=& \frac{1}{R_{\rm A}} +
j  \, \omega C_{\rm A}
\end{eqnarray}
with
\begin{eqnarray}
\label{conductance_array} G_{\rm A}(\omega) &=&\frac{N}{R_{\rm spad}} \\
\label{susceptance_array} B_{\rm A}(\omega)  &=& \omega N\, C_{\rm spad},
\end{eqnarray}
where $C_{\rm spad}$ and $R_{\rm spad}$ are obtained from Eqs.~(\ref{capacitance_spad},~\ref{resistence_spad}), respectively; finally, N\,=\,$64 \times 64=4096$ for the present device.
\par
In Fig.~\ref{cv2} (\ref{Array_R}), the measured  values of the capacitance (resistance) are shown as a function of test frequency up to 1\,MHz, while the continuous line is that obtained using Eq.~(\ref{susceptance_array}) [Eq.~(\ref{conductance_array})].~These measurements were carried out at room temperature using the LCR meter with the SiPM device operated at a reverse bias voltage of 5\,V.~Furthermore, in Table~\ref{Array:fit} the values of $C_{\rm d}$, $R_{\rm d}$, $C_{\rm b}$, $R_{\rm b}$ and $R_{\rm s}$ obtained from such a fit of the SIJD model (adapted to a SiPM) to experimental data
are reported.~It can be remarked that the value obtained for $R_{\rm s}$ is in agreement with one of those typically used in the current technology.
%%%%%%%%%
\begin{table}[t]
 \vskip -.3cm
\begin{center}
\tbl{$C_{\rm d}$, $R_{\rm d}$, $C_{\rm b}$, $R_{\rm b}$ and $R_{\rm s}$ obtained from a fit of the SIJD model (adapted to a SiPM) to experimental data with the device operated at a reverse bias voltage of 5\,V.}
{\begin{tabular}{rl|ccccc}
\toprule
$V_{\rm r}$ [V]& & {$C_{\rm d}$ [pF]}  & {$R_{\rm d}$ [G$\Omega$]}  & {$C_{\rm b}$ [pF]}  &{$R_{\rm b} $[M$\Omega$]} &{$R_{\rm s} $[k$\Omega$]} \\
\colrule
5 & & 0.262 & 385 & 0.056  & 12.5 & 195 \\
\botrule
\end{tabular}}
\label{Array:fit}
\end{center}
\end{table}
%%%%%%%%%
\par
Finally, one can point out that the frequency dependence of the present SIJD model\footnote[6]{The reader can found modelizations of SiPM devices adapted for a time based readout\cite{ST_2} or physical parameters measured at fixed frequencies (e.g., at 100\,kHz in Ref.\cite{Condorelli} and 1\,MHz in Ref.\cite{ST_1}).} for SiPM and photodiode devices is well in agreement with measurements.
\vspace{-.2cm}
\section{Conclusions}
\label{SiPM_Conclusion}
The electrical characteristics of SiPM devices were investigated as a function of applied reverse bias voltage ($V_{\rm r}$) and temperature from 410 down to 100\,K.~The capacitance response was also studied as a function of test frequency.~One finds that the measured capacitance decreases i) with increasing the test frequency above 10\,kHz at 300\,K and ii) with the decreasing of the temperature.~Furthermore, the measurement of the leakage current allowed one to determine the value (and its temperature dependence) of the threshold voltage ($V_{\rm th}$) above which a SiPM begins to operate in Geiger mode: $V_{\rm th}$ decreases with lowering temperature at the rate of $\approx - 29 \,$mV/K from 300 down to 100\,K.
\par
It was developed an electrical model to treat the frequency dependence of a SiPM device operated under reverse bias voltage.~This model is based on the so-called small-signal ac impedance of junction diode (SIJD) used for radiation detectors and photodiodes.~The model was found to be well suited to account for the SiPM dependencies of capacitance and resistance on frequency.
%%%%%%%%%%%%%%%%%%%%%%%%%%%%%%%%%%%%%%%%%%%%%%%%%%%%%%%%%%%%%%%%%%%%%%%%%%%

\end{document}